\documentclass[fleqn,usenatbib]{mnras}

\usepackage{newtxtext,newtxmath}

\usepackage{float} 
\usepackage[T1]{fontenc}

\DeclareRobustCommand{\VAN}[3]{#2}
\let\VANthebibliography\thebibliography
\def\thebibliography{\DeclareRobustCommand{\VAN}[3]{##3}\VANthebibliography}

\newcommand{\xs}{x^{*}}
\newcommand{\ys}{y^{*}}
\newcommand{\zs}{z^{*}}
\newcommand{\DL}{D_{\rm L}}

\DeclareUnicodeCharacter{2212}{-}
\renewcommand{\arraystretch}{1.2}

\usepackage{graphicx}	
\usepackage{amsmath}	
\usepackage{multirow}

\title{Unbiased Bayesian Inference of Peculiar Motions of Galaxies from Type Ia Supernovae Observations}

\author[]{Ujjwal Upadhyay,$^{1,2}$\thanks{E-mail: ujjwalu@iisc.ac.in}
Tarun Deep Saini,$^{1}$
and Shiv K. Sethi$^{2}$
\\
$^{1}$Department of Physics, Indian Institute of Science,
C. V. Raman Road, Bangalore 560012, India \\
$^{2}$Astronomy \& Astrophysics Group, Raman Research Institute,
C. V. Raman Avenue, Bangalore 560080, India
}

\date{Accepted XXX. Received YYY; in original form ZZZ}

\pubyear{\the\year{}}

\begin{document}
\label{firstpage}
\pagerange{\pageref{firstpage}--\pageref{lastpage}}
\maketitle

\begin{abstract}
The peculiar motions of galaxies are powerful cosmological probes that trace the growth of structures and the distribution of matter in the universe, providing a means to investigate the nature of dark energy and test gravity on cosmological scales. However, their direct observation is extremely challenging, as it requires independent and precise distance measurements to galaxies. We present a Bayesian approach to estimate the radial component of peculiar velocities of galaxies hosting Type Ia supernovae (SNe Ia), relying solely on the background cosmological model and the precision of the SNe Ia data. Unlike other peculiar velocity estimators based on Hubble residuals, our method does not assume local linearity of the magnitude-redshift relation or a fixed cosmology, making it unbiased even for large peculiar velocities and self-consistently avoiding bias due to a wrong cosmology. We validate our method using simulated supernova data with the precision of current and upcoming surveys, and further compare it with the linearized estimator to test its efficacy. We show that our estimator has lower bias than the standard estimator and remains consistent even for larger values of $v_{\rm p}/cz$. We also present a Bayesian derivation for the linearized estimator generalized to include the supernova magnitude covariance.   

\end{abstract}





\section{Introduction} \label{sec:first}

Observations of Type~Ia supernovae (SNe~Ia) have been pivotal in shaping our understanding of the cosmos, particularly the dynamics of the late-time universe. Their discovery provided the first direct evidence for the accelerated expansion of the universe~(\cite{SupernovaCosmologyProject:1998vns, SupernovaSearchTeam:1998fmf}), a phenomenon attributed to dark energy or a cosmological constant. As standardizable candles, SNe~Ia serve as precise distance indicators, enabling measurements of cosmic expansion across a wide redshift range and placing tight constraints on cosmological parameters. Analyses of magnitude--redshift data, for example, have yielded precise estimates of the Hubble constant, such as $H_0 = 73.04 \pm 1.04~\mathrm{km \;s^{-1} Mpc^{-1}}$~\citep{Riess:2021jrx}. Despite the growing number of observed supernovae, the precision of cosmological inference from SNe~Ia remains limited by systematic uncertainties in their standardization and photometric calibration~(\cite{DES:2024hip, Dhawan:2024gqy, Efstathiou:2024xcq, Ginolin:2024lfy}). Nevertheless, current analyses have achieved percent-level precision on key cosmological parameters using SNe~Ia data alone. A significant source of uncertainty arises from the peculiar motions of supernova host galaxies~(\cite{Davis_2011, Vanderveld:2008qu, Carreres:2024rji}) that are deviations from the pure Hubble flow caused by gravitational interactions with large-scale structure~(\cite{1980lssu.book.....P}). These motions perturb the observed redshifts and introduce additional scatter in distance measurements. Understanding the redshift dependence of these effects is crucial not only for refining cosmological constraints from SNe~Ia, but also for probing growth of structures and testing gravity on cosmological scales~(\cite{Huterer:2013xky}).

The peculiar motions arise from the gravitational interaction of overdense regions and consist of a large-scale coherent component and small-scale random motions. From linear perturbation theory, the coherent velocities can be predicted from the density contrast on very large scales~(\cite{1980lssu.book.....P,1995PhR...261..271S, Bernardeau:2001qr}); however, this approximation fails on small, nonlinear scales where the random, virialized motions of individual galaxies dominate~(\cite{1983ApJ...267..465D}). Since gravitational instabilities give rise to both the density contrast and peculiar motions, the latter contain roughly half of the information in the phase space of the matter distribution, especially on non-linear scales~(\cite{1995PhR...261..271S, Howlett:2016urc}). Moreover, while the observed galaxy distribution represents only baryonic matter and must be converted to the total matter density using bias modeling, peculiar velocities directly trace the gravitational potential and are therefore free from such bias~(\cite{1984ApJ...284L...9K, Desjacques:2016bnm}). As a result, independent measurements of peculiar velocities can provide complementary information to galaxy clustering surveys. At quasi-linear and nonlinear scales, peculiar velocities and their statistical measures, such as the mean pairwise velocity of matter, can be computed from state-of-the-art cosmological $N$-body ~(\cite{Springel_2005, Jennings_2012, Okumura_2014}). When combined with higher-order galaxy clustering statistics, these peculiar velocities become powerful probes for constraining modified gravity and dark energy models~(\cite{1987MNRAS.227....1K, Linder_2005, Clifton_2012}). In particular, while many of these models predict similar background cosmological evolutions, they exhibit distinct signatures in perturbations, especially on small scales~(\cite{Koyama_2016}). Since dark energy began to dominate only recently, such features cannot be effectively constrained using the CMB, which primarily probes the early universe and large scales. Therefore, peculiar-velocity measurements provide a unique and essential probe of small-scale physics in the late universe~(\cite{Howlett:2016urc}).

Given their unique advantages as cosmological probes, numerous efforts have been devoted to measuring peculiar velocities with increasing precision. The most common approach for coherent motions is velocity field reconstruction, which infers the peculiar velocity field from the observed galaxy distribution by converting galaxy densities into total matter overdensities and applying a linear integral operator, assuming a fiducial cosmology, to relate density to velocity~(\cite{1994ApJ...421L...1N, Carrick_2015}). However, since the reconstructed velocity field is derived from the galaxy density field under an assumed cosmology, it does not introduce additional independent degrees of freedom; rather, it reflects the same information encoded in the density field. Another approach uses empirical relations such as the Fundamental Plane for elliptical galaxies~(\cite{1987ApJ...313...59D, 1987ApJ...313...42D}) and the Tully--Fisher relation for spirals~(\cite{1977A&A....54..661T}) to infer redshift-independent distances and subtract the Hubble flow; while individual measurements are noisy, their correlations ensembled over many galaxies yield cosmological insights. SNe Ia offer a more precise probe, as residuals from fitting the magnitude–redshift relation in a given cosmological model can be translated into peculiar velocities via standard error propagation~(\cite{Hui_2006,  Gordon_2007, Huterer_2017}). The standard Hubble residual-based peculiar velocity estimators are derived under the assumption of local linearity of the magnitude--redshift relation (or distance--redshift relation), which is valid only when the peculiar velocities are very small compared to the Hubble flow ($v_{\rm p}<<cz_{\rm cos}$)~(\cite{Davis_2011, Carreres_2023}). Moreover, the estimator has to be evaluated at the unknown cosmological redshift, $z_{\rm cos}$, and with the true cosmological parameters. In practice, the estimator is evaluated at the observed redshift, $z_{\rm obs}$, and with the fiducial values of the cosmological parameters. For small peculiar velocities, the first approximation is valid; however, the choice of a wrong cosmology can significantly bias the velocity estimates. For $v_{\rm p}\sim cz$, the linear approximation breaks, and the standard estimator becomes biased even for the correct cosmology~(\cite{Carreres_2023}) and therefore cannot be used reliably.

In this work, we present a Bayesian approach to estimating the line-of-sight peculiar velocities of host galaxies using SNe Ia observations. We employ a method developed in a previous work~(\cite{Upadhyay:2025oit}) for fitting general non-linear models with errors in both dependent and independent variables. Using the magnitude–redshift relation in the standard $\Lambda$CDM model without fixing its parameters, we estimate the peculiar velocities at the redshifts of the observed SNe Ia. The method is validated with simulated data having the precision of current and upcoming supernova surveys. We also present a generalized Bayesian version of the standard linear-approximation-based peculiar velocity estimator that accounts for the covariance of SNe Ia magnitudes. Using simulated data, we show that our approach yields consistent and unbiased estimates even for larger peculiar velocities, where the linear approximation breaks down. Further, it self-consistently avoids bias due to incorrect cosmology and is free of the assumption of Gaussianity of the peculiar velocity distribution and therefore works for both random and coherent peculiar motion. 

This paper is organized as follows. In Section~\ref{sec:PV}, we briefly discuss the linear theory of peculiar velocities. In Section~\ref{sec:second}, we discuss the methodology. First, formulation of the Bayesian method for fitting general errors-in-variables models and its specialization for the magnitude--redshift relation in subsection \ref{sec:IIA}, and then the linear approximation-based estimator generalised to account for SNIa magnitude covariance in subsection \ref{sec:IIB}. In Section~\ref{sec:third}, we present the results of testing and validating our methods using simulated supernovae data and their application to the Pantheon+ sample of SNe Ia. In Section~\ref{sec:fourth}, we conclude with a summary of the work discussing the significance and limitations of our approach compared to the existing literature on this problem of peculiar velocity estimation. In 
Appendix \ref{sec:app-A}, we present a Bayesian derivation of the linear approximation-based estimator generalised to account for the supernova magnitude covariance. In Appendix \ref{sec:app-B}, we present trace plots of the velocity estimates at different redshifts and discuss the convergence issues arising from the sensitivity of the likelihood to small redshift perturbations. 
\section{Peculiar Velocity in Linear Perturbations}
\label{sec:PV}
In cosmological theory, peculiar velocities of galaxies arise
from density perturbations. In linear theory, the curl-free velocity
field can be related to the density contrast, in Fourier space,  as~(\cite{1980lssu.book.....P}):
\begin{equation}
  {\mathbf v}({\mathbf k}) =  {i a(t) {\mathbf k} \over k^2} {\partial  \delta({\mathbf k},t) \over \partial t}
\end{equation}
In linear theory, the spatial and time dependence of the density contrast are separable: $ \delta({\mathbf k},t) = D^+(t) \delta({\mathbf  k})$, where $D^+(t)$ is the growing mode of density contrast.  this gives us:
  \begin{equation}
    {\mathbf v}({\mathbf k}) =  {i a(t) {\mathbf k} \over k^2} {d D^+(t)\over dt}\delta({\mathbf k})
    \label{eq:veldenrel}
  \end{equation}
  The peculiar velocities of galaxies cannot be computed  using only  the  linear theory as non-linear effects become important at the scale of galaxies. The peculiar velocity field of galaxies is  modelled as the sum of a large-scale coherent velocity field  and a random, uncorrelated component owing to small scale
  effects. The large-scale, coherent velocity field can be simulated 
  using Eq.~(\ref{eq:veldenrel}) for Gaussian density perturbations, an observed
  feature of linear perturbations.

  We simulate the coherent, large-scale velocity field. 
  These simulations yield the linear component of
  the line-of-sight velocity field.  The details of the simulation are discussed in the results section. 
\section{Methodology} \label{sec:second}
The standard likelihood for estimating cosmological parameters from magnitude--redshift data is sensitive to the contribution of the peculiar velocity of supernova host galaxies to the observed redshifts, especially at low redshifts. Therefore, if not properly accounted for, these peculiar velocities can introduce bias into parameter estimation. Due to this dependence, however, it also presents an opportunity to estimate these peculiar velocities jointly with the cosmological parameters. In order to estimate the peculiar motion of supernovae host galaxies, we treat the fitting of the magnitude--redshift relation to the SNe Ia data as an \emph{errors-in-variables} model with the peculiar motions contributing to the errors in the independent variable, $z$. For this, we use a Bayesian estimator developed in our previous work~(\cite{Upadhyay:2025oit}) to study and correct for the effect of peculiar motion of galaxies on the inference of cosmological parameters from SNe Ia data.
We also discuss a Bayesian derivation of the standard peculiar velocity estimator based on a local linear approximation of the magnitude–redshift relation, assuming Gaussian redshift uncertainties induced by peculiar velocities, and generalize it to incorporate supernova magnitude covariance~(See Appendix \ref{sec:app-A} for the details). Later in the manuscript, we validate and compare these estimators using simulated SNIa data with magnitude uncertainties corresponding to the current and upcoming supernovae surveys. 
\subsection{General MCMC Method}
\label{sec:IIA}
Here, we briefly outline the MCMC-based general method for fitting \emph{errors-in-variables} models and its specific cosmological application to fitting the magnitude--redshift relation to supernova observations as described in detail in~(\cite{Upadhyay:2025oit}), and then discuss its application in peculiar velocity inference, which is the primary focus of the present study. 
Consider a general model expressed in functional form,
\begin{equation}
y_i = f(\boldsymbol{\theta}, x_i).
\end{equation}
where $\boldsymbol{\theta}=\{\theta_1,\theta_2,...\theta_m\}$ are the model parameters, and $x$ and $y$ are the independent and dependent variables of the model, respectively. In the usual scenario where only the dependent variable, $y$, has noisy measurements or the uncertainty in the independent variable is negligible, the likelihood is derived from the distribution of errors in the dependent variable. However, in reality, all measurements are subject to uncertainties, and we often encounter model-fitting problems in which the errors in both variables must be taken into account. Such model-fitting problems are called \emph{errors-in-variables} models~(\cite{dougherty2007introduction}). For the simple linear model and Gaussian distribution of the errors, the problem can be easily solved by analytically marginalizing over the uncertainty in the independent variable to obtain a likelihood based on errors on both dependent and independent variables (\cite{Press:1992:NRC}). For an arbitrary non-linear model with \emph{errors-in-variables}, our approach is to treat the errors in the independent variable as parameters while treating the errors in the dependent variables as usual.
In the usual case with errors only in the dependent variable, we proceed to write the likelihood function assuming that the model predicts the true values, $\ys_i$ and the measurements acquire some random noise, $\epsilon_{Y_i}$, yielding  $y_i= \ys_i +\epsilon_{Yi}$. If the random noise in the measurement has a probability distribution $P_{\rm Y} (\epsilon_Y)$, the likelihood function is given by

\begin{equation}
     \mathcal{L}_Y( \mathcal{D}\mid \boldsymbol{\theta}, \ys) = \prod_i  P_{Y}( y_i- \ys_i)\delta(\ys_i -f( \boldsymbol{\theta},x_i )),
   \label{eq:nonlin0}  
\end{equation}
where $\mathcal{D}$ represents the data, and the Dirac delta function ensures that the true values of the dependent variable $y$ depend on $\boldsymbol{\theta}$ and the measured independent variable $x$ through the model function $f(\boldsymbol{\theta},x)$.\\

\noindent
Using Bayes' theorem, the posterior distribution is given by
\begin{equation}
    P(\boldsymbol{\theta}, \ys \mid \mathcal{D};I) \propto  \mathcal{L}_Y( \mathcal{D}\mid \boldsymbol{\theta}, \ys) P(\boldsymbol{\theta} \mid I) 
\end{equation}
where, $P(\boldsymbol{\theta} \mid I)$ is the prior distribution for the model parameters $\boldsymbol{\theta}$, and $I$ denotes any other background information. When we have errors in both dependent and independent variables, the likelihood in Eq. (\ref{eq:nonlin0}) can be generalized to   
\begin{equation}
     \mathcal{L}_{XY}( \mathcal{D}\mid \boldsymbol{\theta}, \xs,\ys) = \prod_i  P_{Y}( y_i- \ys_i) P_{X}( x_i- \xs_i)\delta(\ys_i -f( \boldsymbol{\theta},\xs_i )),
   \label{eq:jointlikelihood}  
\end{equation}
where $\xs_i$ are the true values of the independent variables, which we treat as parameters, and $P_X$ is the probability distribution of errors in the independent variable, $x_i=\xs_i+\epsilon_{X_i}$. Again, using Bayes' theorem, the posterior distribution for the model parameters $\theta$ and the independent variables $\xs$ can be written as 
\begin{equation}
\label{eq:nonlin2}
    P(\boldsymbol{\theta},\xs \mid \mathcal{D}; I) =  \prod_i  P_Y(y_i-f(\boldsymbol{\theta},\xs_i)) P_{X}( x_i- \xs_i)P({\boldsymbol{\theta}}\mid I) P_{\chi^2} (\chi_Y^2)\,,
\end{equation}
where $P_{X}$ contains our prior information about the independent variable, and $ P_{\chi^2}$ is a $\chi^2-$distribution in $\chi_Y^2$ given by  
\begin{equation}
     P_{\chi^2} (\chi^2_Y) = \frac{\chi_Y^{k-2} \exp(-\chi^2_Y/2)}{2^{k/2}\Gamma(k/2)},
\end{equation}
with $k$ as the desired degrees of freedom of the fit. Here, $P_{\chi^2}$ is a regularization factor, often introduced to mitigate overfitting (\cite{Sivia1996}) -- a common issue in high-dimensional problems such as this, where the number of parameters exceeds the number of data points. Eq. (\ref{eq:nonlin2}) allows us to estimate the parameters $\theta$ along with the correction in the independent variable $x$. In this work, we are interested in fitting the magnitude--redshift relation in the standard $\Lambda$CDM model, where $\theta$ corresponds to the cosmological parameters and the independent variable is the redshift. Eq. (\ref{eq:nonlin2}) thus enables joint estimation of cosmological parameters and redshift uncertainties, which we assume to originate from peculiar velocities.

The relation between the apparent magnitude ($m$) and redshift ($z$) of a source is given by
\begin{equation}
\label{eq:lumdist}
  m=M+  5\log\left(\frac{\DL}{\text{Mpc}}\right) +25\;,  
\end{equation}
where $M$ is the absolute magnitude, and $\DL$ is the luminosity distance of the source. For a flat Friedmann–Lemaître–Robertson–Walker universe, the luminosity distance of a source at redshift $z$ can be expressed as
\begin{equation}
    \DL = \frac{c(1+z)}{H_0}\int_0^z\frac{dz'}{h(z')} ,
\end{equation}
where the dimensionless Hubble parameter is given by
\begin{equation}
    h(z) = \left [\Omega_{\rm M}(1+z)^3 +\Omega_{\rm DE}(1+z)^{3(1+w)}\right ]^{1/2}\,.
\end{equation}
Here $\Omega_{\rm M}$ and $\Omega_{\rm DE}$ are the present-day matter and dark energy density parameters, respectively, and $w$ is the dark energy equation of state parameter.

In this context, the general expression in
Eq. (\ref{eq:nonlin2}) takes the following form:
\begin{equation}
\label{eq:nonlin3}
    P(\boldsymbol{\theta},\zs \mid \mathcal{D}; I) \propto   \exp \left[-\frac{\chi^2_m ( \boldsymbol{\theta}, \zs)}{2}  \right] P_{Z}( z_i- \zs_i)P(\boldsymbol{\theta}\mid I) P_{\chi^2} (\chi_m^2)\,,
\end{equation}
\noindent
where, $\chi^2_m$ given by
\begin{equation} 
\label{eq:chi_m} 
\chi^2_m(\boldsymbol{\theta}, \mathbf{z}) 
= \mathbf{m}^{\mathrm T} \mathbf{C}^{-1}_{m} \mathbf{m},
\end{equation}
with the magnitude residual vector and covariance defined by
\begin{equation}
\mathbf{m} =
\begin{pmatrix}
m_1 - m(z_1^*) \\
\vdots \\
m_N - m(z_N^*)
\end{pmatrix}, \text{and} \; 
\mathbf{C}_m =
\begin{pmatrix}
\sigma_{m_1}^2 & \cdots & \sigma_{m_1 m_N} \\
\vdots & \ddots & \vdots \\
\sigma_{m_N m_1} & \cdots & \sigma_{m_N}^2
\end{pmatrix}. 
\end{equation}\\
For estimating the peculiar velocities, we sample from the above posterior (Eq. (\ref{eq:nonlin3})) by varying both the $\Lambda$CDM parameters, $\boldsymbol{\theta}=\{H_0,\Omega_{\rm M}\}$, as well as the true redshifts $z^*_{i}$. Thus, we obtain the posterior distribution of redshift errors, $\Delta z_i\equiv z_i-\zs_i$, at each redshift in the data, which can be converted into peculiar velocity using the relation, $\Delta z=({v_p}/{c})(1+z)$. Since our method does not require fixing cosmological parameters, it remains free from biases associated with an assumed incorrect cosmology. Moreover, by exploring the full posterior of each redshift parameter, it naturally accounts for correlations among the peculiar velocities. It is important to note that this method can capture only the line-of-sight component of the peculiar velocities. Although this approach provides an unbiased estimate of the peculiar velocities of supernovae as discussed in the next section, it involves a variation in the number of parameters that exceeds the number of data points, which can be very large, making it computationally very expensive and statistically too ambitious for any realistic data. We assess the efficacy and robustness of this approach using simulated datasets that mimic both current observations and those anticipated from future surveys. For comparison, we discuss the standard peculiar velocity estimator based on locally linear approximation of the magnitude--redshift relation and the Gaussian assumption of the redshift uncertainty in the next subsection.

\subsection{Linear Approximation}
\label{sec:IIB}
In this section, we present a Bayesian derivation of the standard Hubble residual-based peculiar velocity estimator from Eq. (\ref{eq:nonlin3}), by assuming local linearity of the magnitude--redshift relation and Gaussian distribution for the redshift errors. Apart from being a complementary Bayesian derivation, it also generalizes the standard estimator by including supernovae magnitude covariance.\\ 
Starting with the posterior distribution in Eq. (\ref{eq:nonlin3}) without the regularization factor, substituting $\chi^2_m$ from Eq. (\ref{eq:chi_m}), and assuming that the redshift errors, $\epsilon_{z_i}$, follow a Gaussian distribution, we can write
\begin{equation}
\label{eq:lin1}
 P(\boldsymbol{\theta},\zs \mid \mathcal{D}; I) \propto \exp \left[ -\frac{1}{2} \mathbf{m^{T}} \mathbf{C}_m^{-1} \mathbf{m} \right] 
    \cdot \exp \left[ -\frac{1}{2} \mathbf{z^{T}} \mathbf{C}_z^{-1} \mathbf{z} \right],
\end{equation}
where the redshift residual vector and associated covariance matrix are given by 
\begin{equation}
\label{eq:lin2}
\mathbf{z} = 
\begin{pmatrix}
z_1 - z_1^* \\
\vdots \\
z_N - z_N^*
\end{pmatrix}, \text{and} \;
\mathbf{C}_z =
\begin{pmatrix}
\sigma_{z_1}^2 & \cdots & \sigma_{z_1 z_N} \\
\vdots & \ddots & \vdots \\
\sigma_{z_N z_1} & \cdots & \sigma_{z_N}^2
\end{pmatrix}. 
\end{equation}
Both these matrices $\mathbf{z}$ and $\mathbf{C}_z$ are unknowns of the problem, since we do not know the true cosmological redshifts $\zs_i$ or the correlation among them. Therefore, these matrices are chosen such that the second factor in Eq.~(\ref{eq:lin1}) represents the prior probability distribution. For example, we can take $\mathbf{C}_z=diag[\sigma_{z},...,\sigma_z]$ with $\sigma_z =0.001$ corresponding to the independently observed peculiar velocities, $v_{\rm p}\sim 250 \; \rm km\;s^{-1}$.

Here, regularization is not required since we are deriving an analytic expression for the redshift errors. In the MCMC analysis, by contrast, the goal was to sample from the distribution, where regularization could help stabilize parameter exploration and prevent overfitting. Further, since we are interested in estimating the redshift parameters (or peculiar velocities), we fix the cosmological parameters at their best-fit values $\theta=\theta_0$ to obtain
\begin{equation}
\label{eq:lin3}
     P(\zs \mid \mathcal{D};\theta_0; I) \propto \exp \left[ -\frac{1}{2} \mathbf{m}^{T} \mathbf{C}_m^{-1} \mathbf{m} \right] 
    \cdot \exp \left[ -\frac{1}{2} \mathbf{z^{T}} C_z^{-1} \mathbf{z} \right],
\end{equation}
Taylor expanding the apparent magnitude about the observed redshift, $z$, and keeping terms upto linear order in ($\zs-z$), we obtain 
\begin{equation}
 \label{eq:lin4}
    m(z^*)=m(z) + m'(z) (z^*-z),
\end{equation}
where $m'(z)$ is the derivative of $m(z^*)$ w.r.t. $z^*$ evaluated at $z^*=z$.
\vskip 5pt \noindent
Substituting (\ref{eq:lin4}) into (\ref{eq:lin3}) and rearranging the terms, we obtain a multivariate Gaussian in ${z^*}$ with mean and covariance given by 
\begin{equation}
 \label{eq:lin5}
    \mathbf{\bar{z}^*} = -\left[ \mathbf{C}_z^{-1} + \mathbf{M'^{T}} \mathbf{C}_m^{-1} \mathbf{M'} \right]^{-1} \left[ \mathbf{M'^{T}} \mathbf{C}_m^{-1} \mathbf{m^o} + \mathbf{C}_z^{-1} \mathbf{z^o} \right],
\end{equation}
and
\begin{equation}
     \label{eq:lin6}
     \mathbf{\Sigma}_{z{^*}} \;=\; \left[\mathbf{C}_z^{-1} + {\mathbf{M}'^{T}} \mathbf{C}_m^{-1}\mathbf{M'}\right]^{-1},
\end{equation}
respectively. Here, $\mathbf{M'}=\text{diag}[m'_1,m'_2,...,m'_{_N}]$, while $\mathbf{m^o} \rm \; and \;\mathbf{z^o}$ are column vectors given by
\begin{equation}
 \label{eq:lin7}
\mathbf{m^o}=
    \begin{pmatrix}
m_1 - m(z_1) + m'(z_1) z_1 \\
\vdots \\
m_{_N} - m(z_{_N}) + m'(z_{_N}) z_{_N}
\end{pmatrix}_{N \times 1}, 
\; \text{and} \; \mathbf{z^o}=
\begin{pmatrix}
z_1 \\
\vdots \\
z_{_N}
\end{pmatrix}. 
\end{equation}
This provides estimates of the true cosmological redshifts and associated uncertainties in their estimation. Subtracting the observed redshift from Eq.~(\ref{eq:lin5}) and converting the difference $\Delta z$ to velocities using the relation $\Delta z=({v_{\rm p}}/{c})(1+z)$, we get an estimate of the peculiar velocities, under the assumption of locally linear approximation of the magnitude--redshift relation and the assumption of Gaussian distribution for the redshift errors. This is a Bayesian derivation of the standard peculiar velocity estimator used in the literature (see Refs. [1, 2, 3]) expressed in matrix form. The main advantage is the generalization to include supernova magnitude covariance and the explicit appearance of the prior through the redshift covariance $C_z$.

The linear approximation holds for small peculiar velocities ($v_{\rm p} << cz$), which occur at high redshifts where the Hubble flow dominates. At low redshifts, $v_{\rm p}\sim cz$, and the estimator becomes biased. Another source of bias in the standard estimator comes from fixed cosmology. The quantities in Eqs.~(\ref{eq:lin5}) and (\ref{eq:lin6}) are supposed to be computed at the cosmological redshifts ($z_{\rm cos}$) and with the correct values of cosmological parameters. However, $z_{\rm cos}$ is an unknown of the problem, and the inferred cosmological parameters themselves could be biased due to the effect of unknown peculiar velocities. In practice, the estimator is evaluated at the fiducial cosmology with the approximation $z_{\rm cos}\approx z_{\rm obs}$. The MCMC-based approach discussed in the previous subsection is free of both sources of bias, as we show in the results section.\\

Apart from the bias discussed, there is another limitation in its application — the estimates depend on the values of $\sigma_{z_i}$, which is what we want to measure. At this point, it is important to note that the Gaussian distribution in redshift errors can be interpreted as a prior on the true redshift parameters. A very small value of $\sigma_{z_i}$ corresponds to a narrow prior which will not allow the redshift parameters, $\zs$, to take any values far from the observed values, $z$. Similarly, a large value of $\sigma_{z_i}$ corresponding to a wide prior will yield a different estimate of $\sigma_{\zs_i}$ depending on the likelihood based on the precision of the apparent magnitude of the SNIa. This effect can be understood more thoroughly by taking various limits of the Eqs. (\ref{eq:lin5}) and (\ref{eq:lin6}) which we discuss in Appendix \ref{sec:app-A}. It is safer to adopt a wide prior so that the posterior is dominated by the data, not the priors; therefore, we take $\sigma_v\sim 1000 \rm \; km \;s^{-1}$, which is considerably larger than the typically observed values, and yet small enough to provide a meaningful constraint.
\section{Results}
\label{sec:third}
Observations of Type Ia supernovae (SNe Ia), and their establishment as standardizable candles, have played a crucial role in advancing our understanding of the expansion history of the Universe and the nature of dark energy. However, a major source of bias in supernova cosmology arises from the peculiar motions of SNe Ia host galaxies. Since peculiar velocities cannot be directly disentangled from the Hubble flow in observations, they can significantly bias cosmological inference if not properly accounted for. Beyond being a source of systematic uncertainty, peculiar velocities also provide an important cosmological probe through their connection to the growth of structure. Accurate measurement of peculiar velocities is therefore an important problem in cosmology. The standard estimator for peculiar velocities based on SNe Ia data is subject to bias due to simplifying assumptions, such as the local linearity of the magnitude--redshift relation and the assumption of a fixed cosmology. In contrast, our MCMC-based general method avoids both of these sources of bias, providing a self-consistent inference from SNIa data.\\
\begin{figure*}
    \centering
    \includegraphics[width=\textwidth]{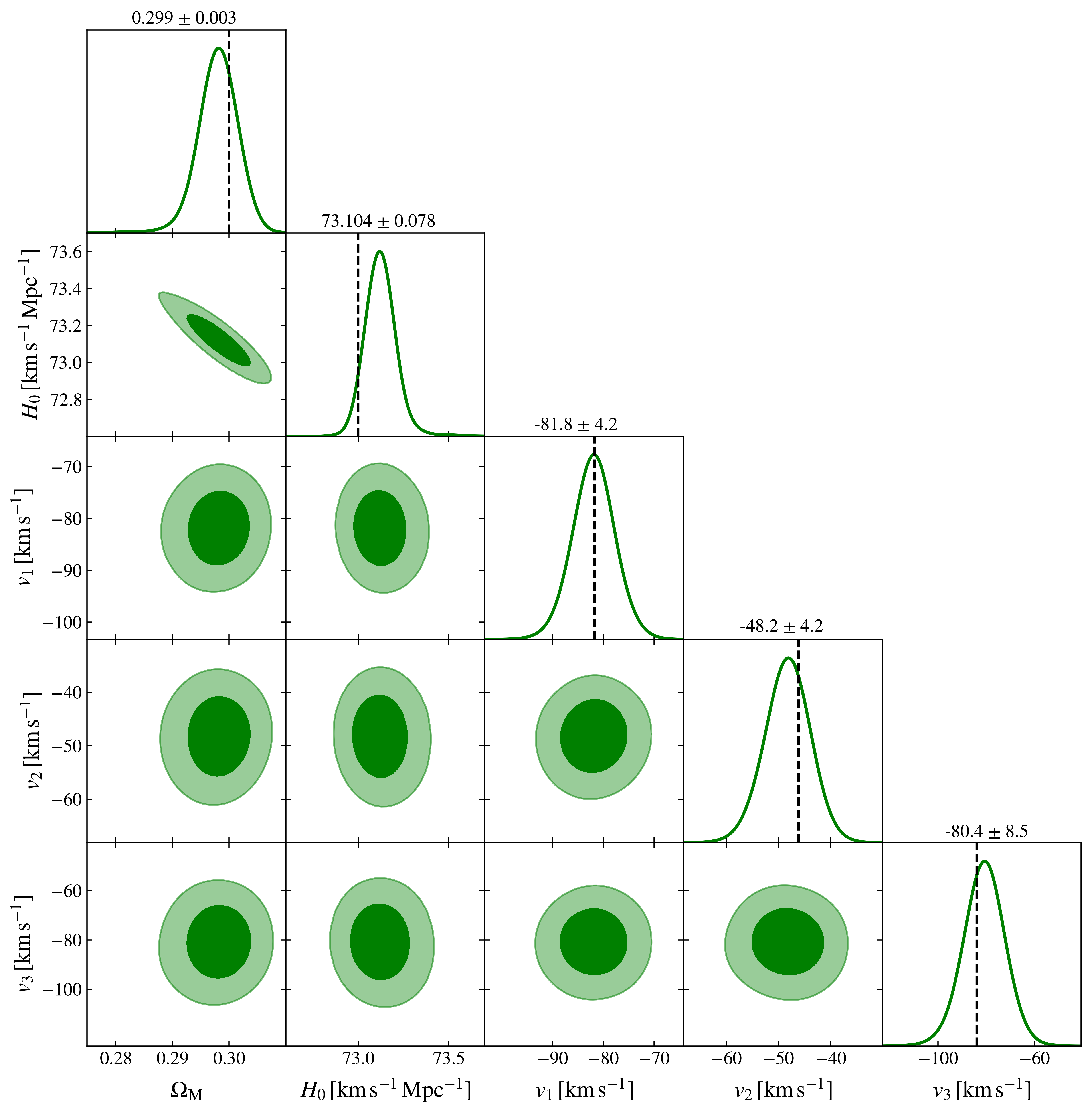}
    \caption{1D and 2D marginalized posterior distribution for the cosmological parameters and peculiar velocities for the three lowest redshift supernovae in the sample obtained from simulated data with $\sigma_m=0.02$. The black-dashed lines show the true values of the parameters.}
    \label{fig:pdf}
\end{figure*}
In this section, we first apply our method to simulated magnitude–redshift data to validate its performance using known input peculiar velocities and cosmological parameters, and to recover their posterior probability distributions. We then compare these results with those obtained using the linearized estimator applied to the same dataset, thereby assessing the performance of the general method relative to the linearized approach. Finally, we apply both methods to the Pantheon+ sample of Type Ia supernovae to investigate the extent to which current data constrain peculiar motions.
\subsection{Validation with Simulated Data}
In order to validate our method, we simulate magnitude--redshift data for 1701 SNIa at the redshifts corresponding to the Pantheon+ sample. For the simulations, we assume the standard $\Lambda$CDM model with fiducial values of the relevant cosmological parameters: $\Omega_{\rm M}=0.30, \Omega_{\Lambda}=0.70$, and $H_0=73 {\rm \; km\;s^{-1}\;Mpc^{-1}}$. The magnitude errors are drawn from a Gaussian distribution, $\Delta m \sim \mathcal{N}(0,\sigma_m)$ for three different cases with $\sigma_m=0.2,0.05$ and $0.02$ that correspond to the current and future precision in the SNIa observations. The effect of random peculiar motion on supernovae cosmology tends to cancel as the number of SNIa in the sample increases, contributing mainly as uncertainty, while the coherent component also introduces significant bias. Therefore, to capture this effect, first we simulate the peculiar velocities on a $6000 \rm \;Mpc$ grid using the velocity power spectrum in the fiducial $\Lambda$CDM cosmology. Essentially, it involves generating a random Gaussian field in real space to represent the matter density contrast, shaping it in Fourier space using the matter power spectrum, computing the peculiar velocity field, and taking an inverse Fourier transform to obtain the real-space velocity field. The real-space velocity field simulated on the grid is then mapped to (RA, DEC, $z$) coordinates so that velocities can be assigned to each SNIa in the sample. After this, we compute the line-of-sight peculiar velocity of the SNIa, convert it to the redshift error, and add it to the observed redshift. In addition to this coherent component, we also add a random motion drawn from a Gaussian distribution $\Delta z \sim \mathcal{N}(0,\sigma_z)$ with $\sigma_z$ corresponding to peculiar velocity, $\sigma_{v_{\rm p}}=250 \; {\rm km/s}$. For the matter power spectrum, we use the Boltzmann code \texttt{CLASS} with the fiducial $\Lambda$CDM parameters.
\subsubsection{General MCMC Method}
First, we apply the general MCMC method to the simulated data to obtain the joint posterior distribution (Eq. \ref{eq:nonlin3}) for cosmological parameters $\theta=\{\Omega_{\rm M}, H_0\}$ and the true redshifts, $\zs$. For this, we need to provide prior information $P(\theta\mid I)$ and $P_Z(z_i-\zs_i)$. We take uniform priors for all the parameters as listed in Table~\ref{tab:prior}.
\begin{table}
\centering
\renewcommand{\arraystretch}{1.9}
    \centering

    \begin{tabular}{cp{1.5cm}c}
    \hline \hline
        \textbf{Parameter} && \textbf{Prior} \\
    \hline 
         $\Omega_{\rm M}$ && $\mathcal{U}(0.1,\,0.5)$ \\
$H_{0}\;[\;\mathrm{km\,s^{-1}\,Mpc^{-1}}]$ && $\mathcal{U}(50,\,80)$ \\
$\zs$ && $\mathcal{U}(-0.003,\,0.003)$ \\
\hline 
    \end{tabular}
    \caption{Priors on the cosmological parameters used in the analysis.}
    \label{tab:prior}
\end{table}
We employ the MCMC sampling technique implemented through the Metropolis-Hastings algorithm in a Python code to sample from the posterior distribution. We run 12 different chains starting randomly from the prior range for the cosmological parameters. The natural starting point for the redshift parameters is $\zs=z$, i.e, assuming that there is no contribution to the observed redshift from the peculiar motion. For the convergence, we use the Gelman--Rubin criterion, $R-1<0.05$ for the cosmological parameters. It is important to mention here that all redshift parameters $\zs_i$ do not converge, marking the limitation of the method and the precision of the data. The convergence depends on the sensitivity of the likelihood to the changes in the parameters, which in turn depends on the curvature of the likelihood and the precision of the data. For the magnitude--redshift relation, the likelihood is sensitive to the redshift errors (or peculiar velocities) at low redshift. A detailed discussion of this convergence issue and its dependence on the redshift is provided in appendix~\ref{sec:app-B}.

Figure \ref{fig:pdf} shows the posterior distributions of the cosmological parameters together with the peculiar velocities for the three lowest-redshift SNIa in the simulated sample. The black dashed lines indicate the true values used in the simulation. All five parameters are found to be statistically consistent with their true values, demonstrating the validity of our method. With improved SNIa magnitude precision from future surveys, this approach can enable reliable estimation of peculiar velocity directly from SNIa magnitude–redshift measurements. It is well understood that as we go towards higher redshifts in observations, the effect of peculiar velocity becomes less prominent on cosmological parameters because the peculiar velocity becomes very small relative to the Hubble recessional velocity at those redshifts. However, there is another effect that is more relevant to our discussion---the sensitivity of redshift uncertainty to the apparent magnitude. Due to the logarithmic dependence of the magnitude--redshift relation, the apparent magnitude at high redshift becomes insensitive to small changes in the redshift, which leaves the redshift parameters unconstrained.  

This can be observed in the posterior distribution of peculiar velocities corresponding to high-redshift SNe Ia. However, given the large number of these parameters, it is more convenient to plot the estimated peculiar velocities with their uncertainties at different redshifts rather than the full posterior distributions.

\begin{figure}
    \centering
    \includegraphics[scale=0.48]{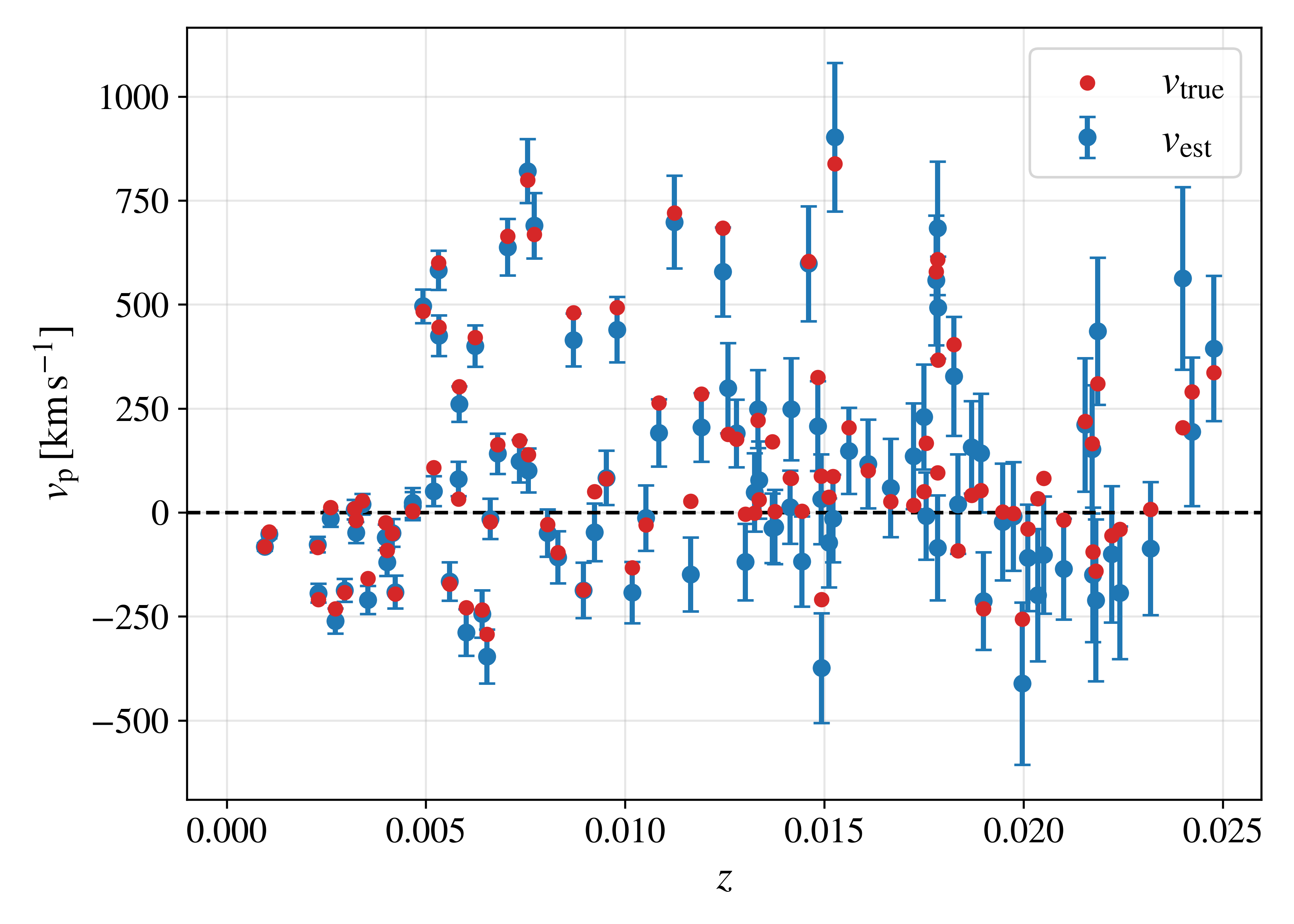}
    \caption{Estimated peculiar velocities along with $1\sigma$ error bars at different redshifts obtained using the general method from simulated data with $\sigma_m = 0.02$.}
    \label{fig:vp_z_sim}
\end{figure}

\begin{figure}
    \centering
    \includegraphics[scale=0.50]{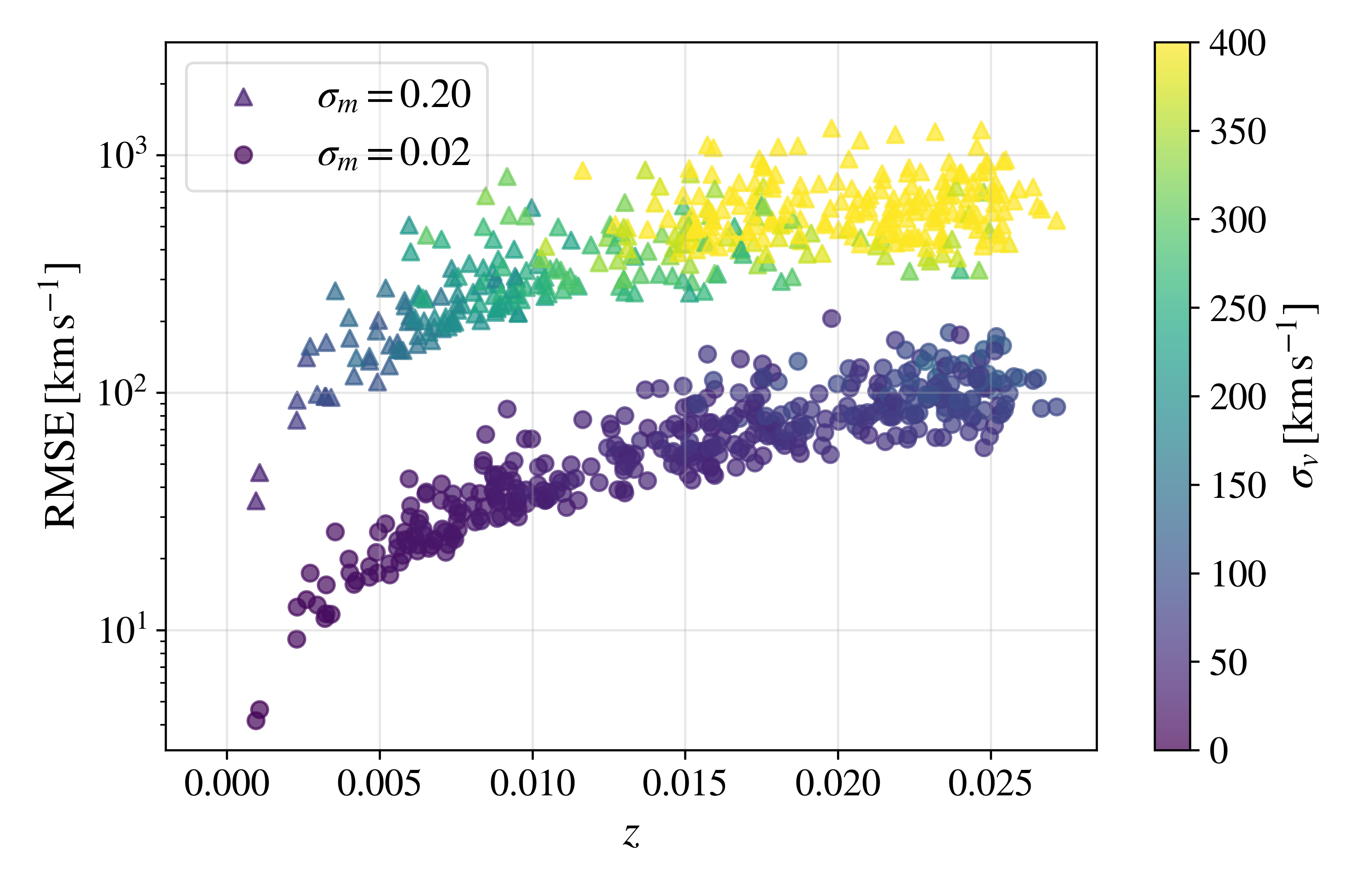}
    \caption{The root mean square errors (RMSE) of estimation as a function of redshifts for two different supernovae magnitude precision shown by different marker shapes. The color bar on the right shows the contribution of variance.}
    \label{fig:RMSE}
\end{figure}

Figure~\ref{fig:vp_z_sim} shows the estimated peculiar velocities as a function of redshift, along with their $1\sigma$ uncertainties. The true velocities are also shown (in black) for comparison. We find that the estimates are statistically consistent with the true values at all redshifts. However, the uncertainties increase toward higher redshifts, and the estimates become effectively prior-dominated, rendering them uninformative. This is the convergence issue mentioned earlier and discussed in detail in Appendix~\ref{sec:app-B}. In contrast, at low redshifts, the peculiar velocities are well constrained and consistent with the true values. The estimates also depend on the precision of the supernova data, as illustrated by the root mean square error (RMSE) in Figure~\ref{fig:RMSE}. Different marker shapes correspond to different levels of magnitude precision, while the color scale indicates the contribution of variance to the RMSE. As expected, the uncertainty in the estimates increases as the precision of the supernova magnitudes decreases. For a fixed precision and redshift, larger values of the RMSE are primarily driven by increased variance rather than bias. This indicates that the overall error budget is dominated by statistical fluctuations, supporting the statistical consistency of the estimator.

\begin{figure}
    \centering
    \includegraphics[scale=0.56]{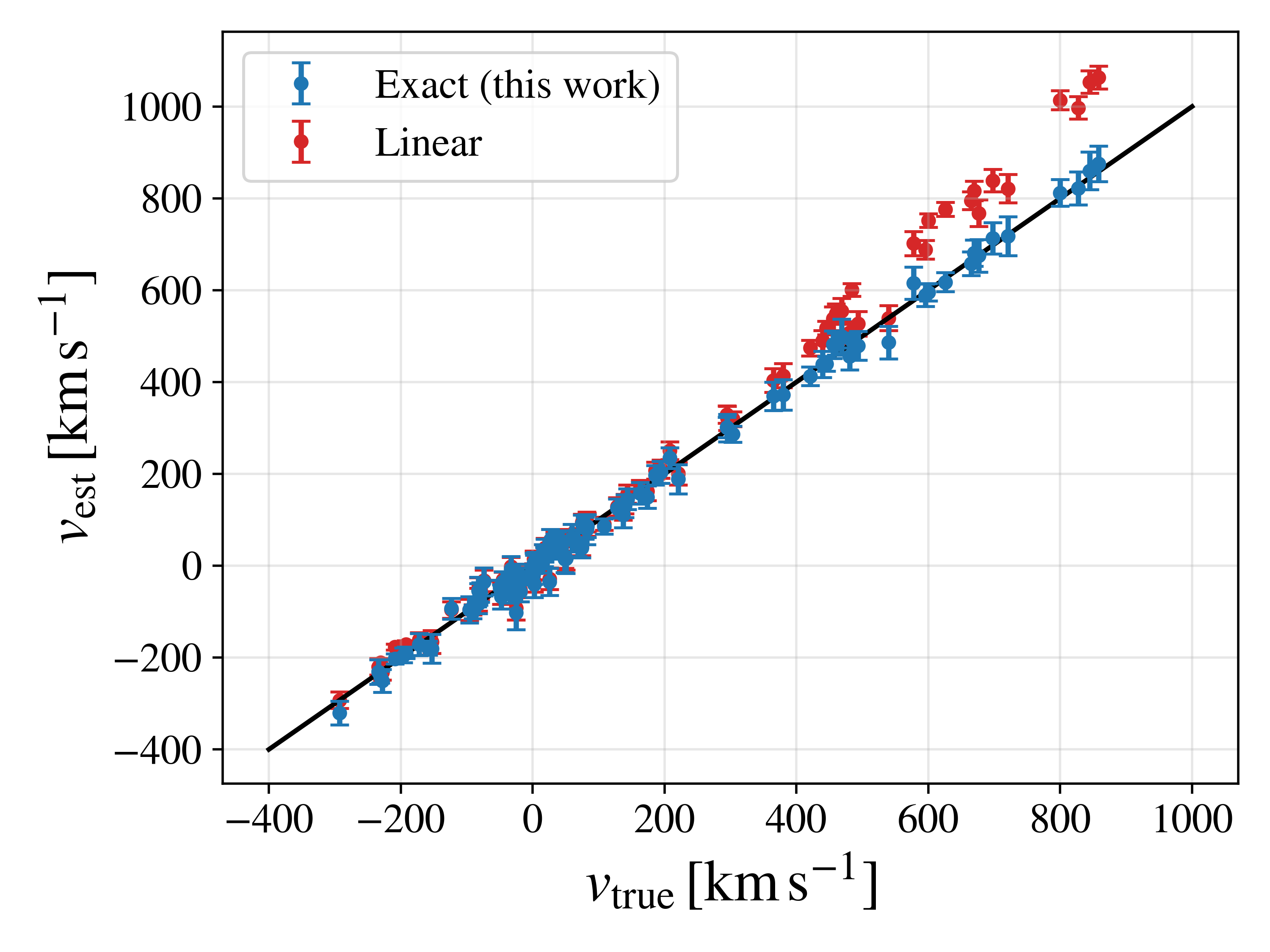}
    \caption{Estimated peculiar velocities along with $1\sigma$ error bars versus the true values used in the simulation for simulated data with $\sigma_m = 0.02$. Different colors represent the methods used.}
    \label{fig:V_est-V_true}
\end{figure}

\begin{figure}
    \centering
    \includegraphics[scale=0.55]{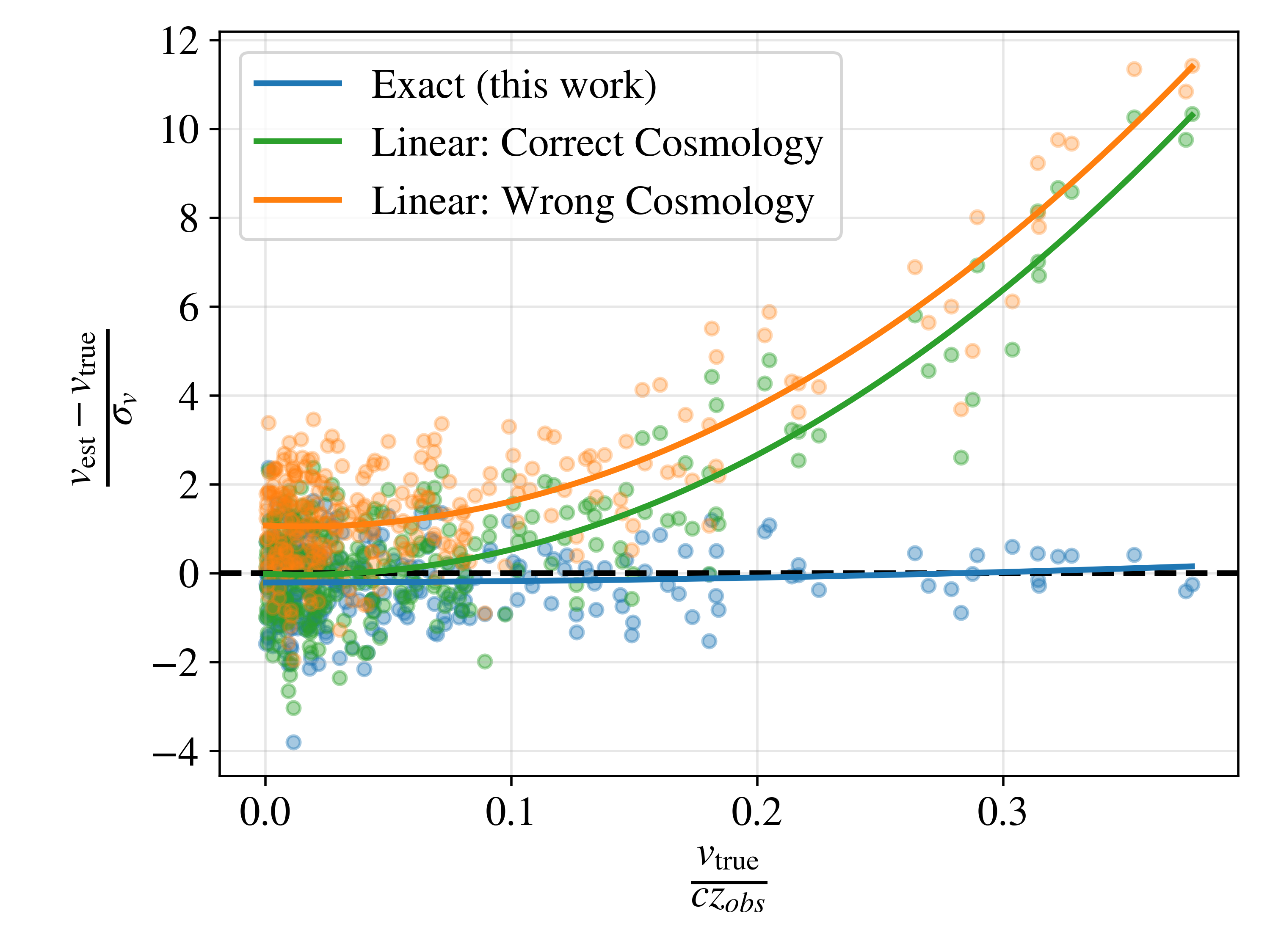}
    \caption{Comparison of bias due to breakdown of linear approximation and wrong cosmology. The solid lines are polynomial fits to the scatter used to better represent the trend. The general method remains statistically unbiased within a $68\%$ credible region, even for $v_{\rm true}\approx cz$, where the linear approximation is not valid.}
    \label{fig:bias}
\end{figure} 
\subsubsection{Linear vs. Exact Method}
Having discussed the results of the general MCMC method above, and demonstrated its effectiveness in estimating peculiar velocities at low redshifts, we now proceed in this subsection to compare the general method with the standard estimator based on linear approximation.
\begin{figure}
    \centering
    \includegraphics[scale=0.55]{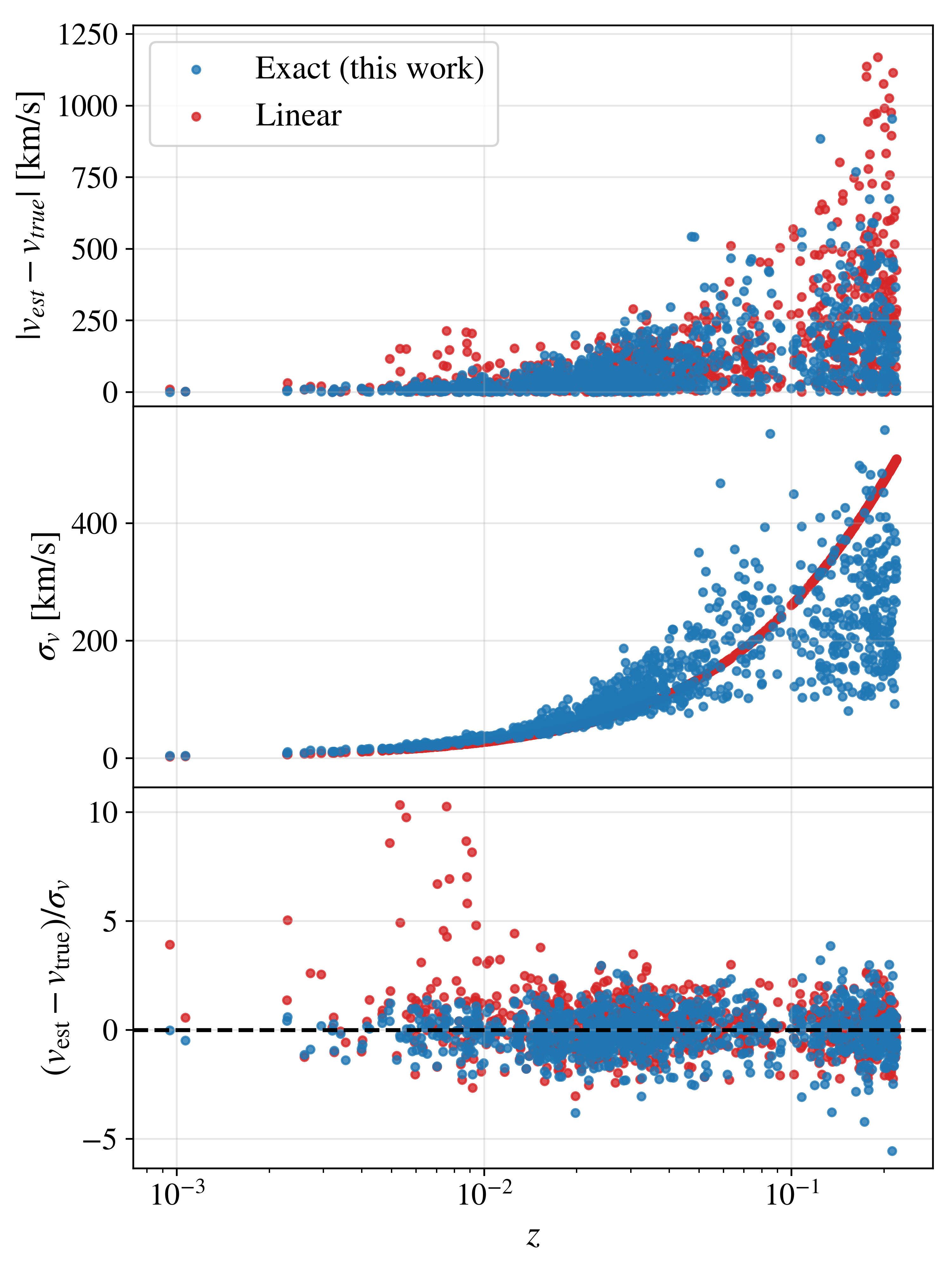}
    \caption{Comparison of the redshift dependence of bias and variance for the linear approximation and the MCMC-based general methods. Both the bias and the variance increase with redshift; however, the estimators remain statistically consistent in most cases. An exception occurs when $v_{\rm true} \approx cz$, where the linear approximation breaks down and the corresponding estimator becomes significantly biased.}
    \label{fig:bias_var}
\end{figure}

Figure~\ref{fig:V_est-V_true} shows the estimated peculiar velocities, with $1\sigma$ uncertainties, plotted against the true velocities used in the simulation for $\sigma_m = 0.02$, comparing the exact (blue) and linear (red) methods. The black solid line corresponds to the ideal relation, $v_{\rm est} = v_{\rm true}$. For small peculiar velocities, the estimates from both the linear and exact methods lie along the black solid line, indicating accurate recovery, as expected since the linear approximation is valid in this regime. However, as $v_{\rm true}$ increases, the linear method begins to overestimate the velocities. In contrast, the MCMC-based general method remains consistent with the true peculiar velocities, although with slightly larger uncertainties than those obtained with the linear method. Despite the larger uncertainty, the general method provides estimates that are closer to the true values, and therefore performs better overall. This is more clearly understood in terms of the bias--variance trade-off: while the linear estimator exhibits smaller variance, it suffers from increasing bias at larger velocities, whereas the general method remains approximately unbiased.

Figure~\ref{fig:bias} shows the normalized bias of the estimator, defined as $(v_{\rm est} - v_{\rm true})/\sigma_v$, plotted as a function of the ratio of the true peculiar velocity to the Hubble recessional velocity, $v_{\rm true}/cz$. The results are shown for the MCMC-based general method (blue), the linear method with the correct cosmology (green), and the linear method assuming an incorrect cosmology (orange). The solid lines are the polynomial fit to the scatter to show the trend. The black dashed line represents no bias. We find that for the linear method, the normalized bias remains small when the peculiar velocities are much smaller than the Hubble recession velocity. However, as the true velocity becomes comparable to the Hubble flow, the bias increases significantly, reaching several standard deviations from zero. The use of an incorrect cosmology, primarily through an incorrect value of $H_0$, introduces an additional systematic bias. In contrast, for the MCMC-based estimator, the normalized bias remains within $\pm 1$, indicating that the estimates are consistent with the true values within the $68\%$ credible interval.

As discussed earlier, the estimates degrade toward higher redshifts because the sensitivity of the likelihood to the small redshift perturbations induced by peculiar velocities decreases with increasing redshift. To illustrate this behaviour, Figure~\ref{fig:bias_var} shows the absolute bias, the standard deviation, and the normalized bias (bias divided by standard deviation) as functions of redshift in three vertical panels, for both the linear and the MCMC-based general methods. For both estimators, the absolute bias and the standard deviation increase with redshift. However, the normalized bias remains approximately constant, indicating that the bias stays within the quoted uncertainties and that the estimators are statistically consistent. Nevertheless, the growing variance at high redshift renders the estimates effectively uninformative, as the posteriors become prior-dominated. For the linear method, we further observe statistically inconsistent estimates even at some low redshifts. These correspond to cases with large values of $v_{\rm true}/cz$, where the linear approximation breaks down. In contrast, the MCMC-based estimator remains consistent in this regime and therefore performs better, particularly at low redshifts where the peculiar velocity is comparable to the Hubble recession velocity.
\subsubsection{Estimating Peculiar Velocity Correlation}
\begin{figure}
    \centering
    \includegraphics[scale=0.48]{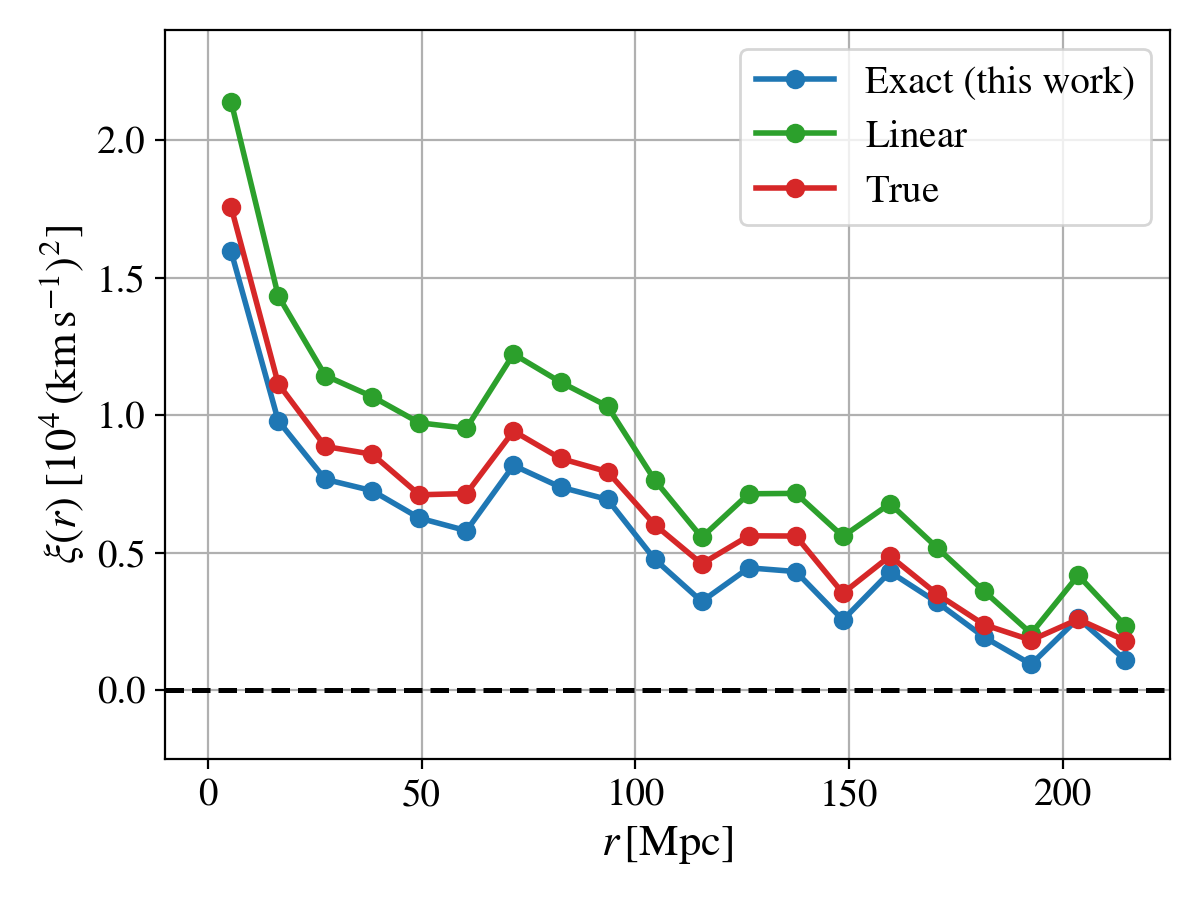}
    \caption{Estimated and true binned two-point correlation function of line-of-sight peculiar velocity for simulated data with $\sigma_m=0.02$.}
    \label{fig:corr}
\end{figure}

The peculiar velocity correlation function encodes information about
the underlying cosmological perturbations. Once measured, it provides a
powerful probe of the growth of structure, tests of gravity on
cosmological scales, and the nature of dark energy. This approach is
particularly promising in light of the large supernova samples expected
from upcoming surveys. However, any bias introduced by the linear
approximation propagates into the inferred correlation function and
can consequently affect the cosmological constraints derived from it. In Figure~\ref{fig:corr}, we compare the velocity correlation functions
obtained using the linear estimator and our MCMC-based method.
However, it is important to note that, given the relatively small
number of supernovae considered here, we have not employed a
state-of-the-art estimator for the correlation function, and the
results should therefore be regarded as illustrative.regarded as illustrative.

A more realistic assessment requires simulations of the peculiar velocity field based on $N$-body calculations, together with supernova
samples constructed to reflect the characteristics of specific surveys.
Such an analysis is necessary to quantify the impact on the inferred
growth of structure and to assess the potential for distinguishing
between dark energy, modified gravity, and interacting dark-sector
scenarios. We leave this detailed investigation for future work.
\subsection{Application to Pantheon+ Data}
In this section, we present the estimates of peculiar velocities derived from the Pantheon+ sample of SNIa using both the linear and exact methods. The Pantheon+ compilation includes 1701 SNIa with spectroscopic redshifts extending up to $z = 2.26$. Supernova surveys measure the apparent magnitudes and redshifts of SNIa; converting luminosity distances into apparent magnitudes requires knowledge of the absolute magnitude, for which we adopt the SH0ES calibration. In the Pantheon+ dataset, redshifts are already corrected for the motion of the solar system relative to the CMB rest frame, as well as for the peculiar motions of host galaxies based on measurements using the Tully–Fisher and Fundamental Plane relations. In our analysis, however, we make use of the CMB-frame redshifts, $z_{\rm cmb}$, since our goal is to estimate these peculiar velocities directly from the supernova data and compare them with existing measurements.           

\begin{figure}
    \centering
    \includegraphics[scale=0.55]{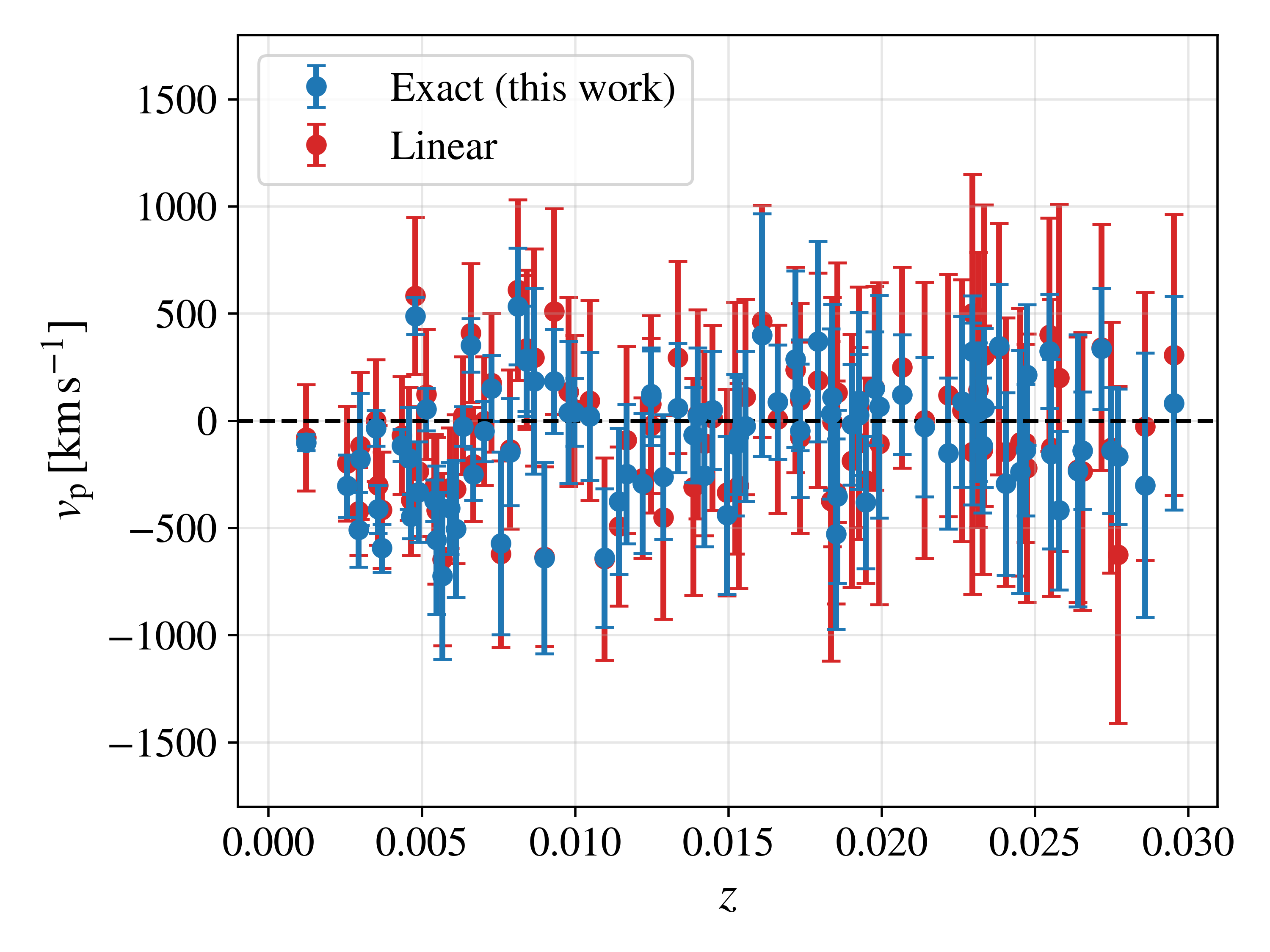}
    \caption{Estimated peculiar velocities along with $1\sigma$ error bars obtained from Pantheon+ SNIa using both linearized and the MCMC-based exact likelihood methods.}
    \label{fig:pan+}
\end{figure}

Figure \ref{fig:pan+} presents the peculiar velocity estimates from the Pantheon+ sample, along with their associated $1\sigma$ uncertainties. At low redshifts ($z < 0.02$), the estimates are consistent with zero, with a standard deviation of $\sigma_{v_{\rm p}} \sim 300 \;\rm km/s$. At higher redshifts, however, the method loses constraining power due to the limited precision of current data and the reduced sensitivity of the likelihood to peculiar velocities. Thus, with present SNIa data, peculiar velocities can only be reliably estimated at very low redshifts. Looking ahead, the much larger samples expected from LSST and ZTF will provide the statistical precision needed to extend these estimates to higher redshifts, as demonstrated with our simulated data. The key strength of our approach lies in its robustness, as it relies on fewer assumptions and remains straightforward to implement compared to galaxy survey–based methods.

\section{Conclusion and Discussion}
\label{sec:fourth}
The peculiar motion of galaxies is an unbiased tracer of matter distribution. This makes it a very promising probe of the growth of structures in the universe. Since peculiar velocities contain information complementary to galaxy clustering, they can be used to investigate the nature of dark energy, test deviations from general relativity on cosmological scales, and distinguish between these two effects. Furthermore, cosmological observations that rely on galaxy redshifts, such as those using standard candles or standard sirens, can be significantly biased if the contribution of galaxy peculiar velocities is not properly accounted for (\cite{ Mukherjee:2019qmm, Nimonkar:2023pyt}). For instance, neglecting peculiar motion of host galaxies can bias our inference of cosmological parameters from the SNIa data by a few percent (\cite{Davis_2011, Upadhyay:2025oit}). This has led to numerous approaches for reconstructing peculiar velocities from observations, encompassing both traditional distance-indicator methods and modern Bayesian or machine-learning-based techniques (\cite{ Branchini:2001fz, Mohayaee:2004he,  Wu:2023wmj, Guachalla:2023lbx, Said:2024pwm}). However, these methods rely on empirical relations, such as the Tully–Fisher and Fundamental Plane relations, which introduce additional assumptions. Combined with the assumptions inherent to galaxy surveys, this makes the inference of peculiar velocities more methodology-dependent and potentially biased. SNe Ia provide another way to estimate peculiar velocities by propagating the Hubble residuals to redshift using standard error propagation. However, the standard Hubble residual-based estimators~\citep{Carreres_2023} rely on a locally linear approximation to the magnitude–redshift relation, which is valid only when the peculiar velocity is small compared to the Hubble flow ($v_p<<cz$). In this work, we present an alternative Bayesian method for estimating the line-of-sight peculiar velocities of SNe~Ia host galaxies from magnitude--redshift data without assuming local linearity. The method
relies solely on the cosmological model and the precision of supernova
observations, making it applicable even at low redshifts where
$v_p \approx cz$ and the linear approximation breaks down.

Our approach to estimating peculiar velocities from supernova data is
based on a method developed for fitting non-linear models with errors
in both dependent and independent variables~\citep{Upadhyay:2025oit}.
In this framework, we fit the magnitude--redshift relation to SNe~Ia
data, treating the apparent magnitudes and redshifts as the dependent
and independent variables, respectively. Following \citep{Upadhyay:2025oit},
and as described in Section~\ref{sec:second}, we treat the redshifts as
free parameters and infer their posterior distributions. The difference
between the observed and inferred redshifts is then attributed to
peculiar motion, which we convert into estimates of peculiar velocities.
To obtain the posterior distribution, we perform MCMC sampling over all
redshift parameters. By exploring the full parameter space, the method
naturally allows us to capture correlations in the velocity field.
We validate the approach using simulated SNe~Ia with a peculiar velocity field generated from a fiducial $\Lambda$CDM cosmology. For comparison, we also derive an analytic expression for the standard
peculiar velocity estimator based on the locally linear approximation of
the magnitude--redshift relation and the assumption of Gaussian
redshift errors, and extend it to include the supernova magnitude
covariance.

The main results of this work are shown in
Figures~\ref{fig:pdf} and~\ref{fig:V_est-V_true}. We find that treating
redshifts as free parameters allows peculiar velocities to be estimated
using the exact likelihood, making the estimator unbiased even when the
peculiar velocity is comparable to the Hubble flow, as illustrated in
Figure~\ref{fig:V_est-V_true}. This is particularly relevant at low
redshifts. In addition, the method provides the full posterior
distribution, which contains additional information about velocity
correlations. However, the standard method based on the linear approximation remains
computationally more efficient and therefore performs better within the
regime where the approximation is valid.

Overall, our method complements existing approaches to estimating
peculiar velocities and can be naturally extended to other distance
indicators, such as standard sirens. Important directions for future
work include studying the applicability of this approach to the larger samples of SNe~Ia expected from upcoming surveys and incorporating
more realistic peculiar-velocity fields from $N$-body simulations,
particularly on non-linear scales, which could help test various models of dark energy
and gravity on cosmological scales, which we leave for future work.
\section*{Acknowledgments} \label{sec:acknowledgments}
UU acknowledges Yashi Tiwari for her valuable comments and suggestions throughout this project. We also acknowledge the use of the HPC cluster at the Raman Research Institute, Bangalore, India.

\section*{Data Availability}

The PantheonPlus sample of Type Ia supernovae data analyzed in this work is publicly available.



\bibliographystyle{mnras}
\bibliography{references} 




\appendix

\section{Derivation of the Linearized Estimator}
\label{sec:app-A}
Here, we provide a detailed derivation of the linearized Bayesian estimator outlined earlier in section \ref{sec:IIB}. Again, starting with the posterior distribution in Eq. (\ref{eq:nonlin3}) without the regularization factor, substituting $\chi^2_m$ from Eq. (\ref{eq:chi_m}), and assuming that the redshift errors, $\epsilon_{z_i}$, follow a Gaussian distribution, we can write
\begin{equation}
\label{eq:lin1}
 P(\theta,\zs \mid \mathcal{D}; I) \propto \exp \left[ -\frac{1}{2} \mathbf{m^{T}} \mathbf{C}_m^{-1} \mathbf{m} \right] 
    \cdot \exp \left[ -\frac{1}{2} \mathbf{z^{T}} \mathbf{C}_z^{-1} \mathbf{z} \right],
\end{equation} 
Taking the logarithm, we can write
\begin{equation}
\ln P(\theta,\zs \mid \mathcal{D}; I) = -\frac{1}{2} \left[ \mathbf{m^{T}} \mathbf{C}_m^{-1} \mathbf{m} 
+  \mathbf{z^{T}} \mathbf{C}_z^{-1} \mathbf{z} \right],
\end{equation}
where the matrices $\mathbf{m}$ and $\mathbf{z}$ represent the supernova magnitude and redshift residual vectors, respectively, and $\mathbf{C}_m$ and $\mathbf{C}_z$ are the associated covariance matrices, defined as:
\begin{equation}
\mathbf{m} = 
\begin{pmatrix}
m_1 - m(\zs_1) \\
\vdots \\
m_{_N} - m(\zs_{_N})
\end{pmatrix}, \text{and} \;
\mathbf{C}_m =
\begin{pmatrix}
\sigma_{m_1}^2 & \cdots & \sigma_{m_1 m_N} \\
\vdots & \ddots & \vdots \\
\sigma_{m_N m_1} & \cdots & \sigma_{m_N}^2
\end{pmatrix}. 
\end{equation}
\begin{equation}
\mathbf{z} = 
\begin{pmatrix}
z_1 - z_1^* \\
\vdots \\
z_N - z_N^*
\end{pmatrix}, \text{and} \;
\mathbf{C}_z =
\begin{pmatrix}
\sigma_{z_1}^2 & \cdots & \sigma_{z_1 z_N} \\
\vdots & \ddots & \vdots \\
\sigma_{z_N z_1} & \cdots & \sigma_{z_N}^2
\end{pmatrix}. 
\end{equation}
Taylor expanding the magnitude--redshift relation about the observed redshift, and keeping terms up to linear order in $z$, we get
\begin{equation}
m(z^*) = m(z) + m'(z)\,(z^* - z).
\end{equation}
Since we attribute the difference between true redshift and observed redshift to the peculiar velocity, the linear approximation breaks when $v_{\rm p}\approx cz$ and higher order terms begin to contribute significantly. Under this approximation, the magnitude residual vector takes the form,
\begin{equation}
\mathbf{m} =
\begin{pmatrix}
m_1 - m(z_1) - m'(z_1)(z_1^* - z_1) \\
\vdots \\
m_N - m(z_N) - m'(z_N)(z_N^* - z_N)
\end{pmatrix}
\end{equation}
For subsequent calculations, it is convenient to decompose the residual vectors into the sum of two components: one that depends solely on the observed quantities, and another that contains the true redshift parameters, $\zs$, as follows. 
Let $\mathbf{m}=\mathbf{m^o}+\mathbf{m^*}$, and $\mathbf{z}=\mathbf{z^o}+\mathbf{z^*}$, with
\begin{equation}
\mathbf{m^o} = 
\begin{pmatrix}
m_1 - m(z_1) + m'(z_1) z_1 \\
\vdots \\
m_N - m(z_{_N}) + m'(z_{_N}) z_{_N}
\end{pmatrix}_{N \times 1}, 
\mathbf{m^*} = -
\begin{pmatrix}
m'(z_1) z_1^* \\
\vdots \\
m'(z_{_N}) z_{_N}^*
\end{pmatrix}_{N \times 1}
\end{equation}
and 
\begin{equation}
    \mathbf{z^o} = 
\begin{pmatrix}
z_1 \\
\vdots \\
z_N
\end{pmatrix}, 
\mathbf{z^*} = -
\begin{pmatrix}
z_1^* \\
\vdots \\
z_N^*
\end{pmatrix}
\end{equation}
With this decomposition of the vectors, the log-posterior can be written as
\begin{align}
\ln P &= -\tfrac{1}{2} \Bigg[
\big(\mathbf{m}^o + \mathbf{m}^*\big)^T \mathbf{C}_m^{-1} \big(\mathbf{m}^o + \mathbf{m}^*\big)
+ \big(\mathbf{z}^o + \mathbf{z}^*\big)^T \mathbf{C}_z^{-1} \big(\mathbf{z}^o + \mathbf{z}^*\big)
\Bigg] \\[0.5em]
&= -\tfrac{1}{2} \Big[
\mathbf{m}^{o T} \mathbf{C}_m^{-1} \mathbf{m}^o + \mathbf{m}^{o T} \mathbf{C}_m^{-1} \mathbf{m}^* 
+ \mathbf{m}^{*T} \mathbf{C}_m^{-1} \mathbf{m}^o + \mathbf{m}^{*T} \mathbf{C}_m^{-1} \mathbf{m}^* \nonumber \\
&\quad + \mathbf{z}^{o T} \mathbf{C}_z^{-1} \mathbf{z}^o + \mathbf{z}^{o T} \mathbf{C}_z^{-1} \mathbf{z}^* 
+ \mathbf{z}^{*T} \mathbf{C}_z^{-1} \mathbf{z}^o + \mathbf{z}^{*T} \mathbf{C}_z^{-1} \mathbf{z}^* 
\Big]
\end{align}
Further, $\mathbf{m^*}$ can be written as $\mathbf{m^*} = \mathbf{M'} \mathbf{z^*}$, where $\mathbf{M'}$ is an $N\times N$ square matrix given by $\mathbf{M'}=\text{diag}[m'_1,...m'_{_N}]$.
Substituting this and completing the square in $\mathbf{z^*}$ (See \cite{10.5555/1162264} for quadratic completion of matrix expressions), we get
\begin{equation}
\label{eq:lin_logpost}
\begin{aligned}
\ln P &= -\frac{1}{2}
\left[
\mathbf{z}^* + \left(\mathbf{C}_z^{-1} + \mathbf{M'}^{T} \mathbf{C}_m^{-1} \mathbf{M'} \right)^{-1}
\left( \mathbf{M'}^{T} \mathbf{C}_m^{-1} \mathbf{m}^o + \mathbf{C}_z^{-1} \mathbf{z}^o \right)
\right]^T \\
& \quad \times
\left( \mathbf{C}_z^{-1} + \mathbf{M'}^{T} \mathbf{C}_m^{-1} \mathbf{M'} \right)\\
& \quad \times
\left[
\mathbf{z}^* + \left(\mathbf{C}_z^{-1} + \mathbf{M'}^{T} \mathbf{C}_m^{-1} \mathbf{M'} \right)^{-1}
\left( \mathbf{M'}^{T} \mathbf{C}_m^{-1} \mathbf{m}^o + \mathbf{C}_z^{-1} \mathbf{z}^o \right)
\right] \\
& \quad + \text{constant},
\end{aligned}
\end{equation}
where the constant denotes the remaining terms that are independent of $\mathbf{Z^*}$.\\
\vskip 2pt
\noindent
Thus, we get a posterior distribution which is a Gaussian in $\mathbf{z^*}$ centered at $\mathbf{\bar{z}^*}$ with the associated covariance $\mathbf{\Sigma}_{\zs}$, given by
\begin{equation}
 \label{eq:lin5}
    \mathbf{\bar{z}^*} = -\left[ \mathbf{C}_z^{-1} + \mathbf{M'}^{T} \mathbf{C}_m^{-1} {M'} \right]^{-1} \left[ \mathbf{M'}^{T} \mathbf{C}_m^{-1} \mathbf{m^o} + \mathbf{C}_z^{-1} \mathbf{z^o} \right],
\end{equation}
and
\begin{equation}
     \label{eq:lin6}
     \mathbf{\Sigma}_{z{^*}} \;=\; \left[\mathbf{C}_z^{-1} + {\mathbf{M'}^{T}} \mathbf{C}_m^{-1} \mathbf{M'}\right]^{-1}.
\end{equation}\\
\noindent
It is instructive to examine various limiting cases of the above expressions in order to better understand the behaviour of the general estimator.\\
\vskip 2pt \noindent
{\bf Case-1:} High $z$\\ Due to the logarithmic form of the magnitude--redshift relation, we have $m'(z) \rightarrow 0$ as $z \rightarrow \infty$. In this limit, $\mathbf{\bar{z}^*} = \mathbf{z^o}$ and $\mathbf{\Sigma}_{z^*} = \mathbf{C}_z$. This implies that the predicted apparent magnitude becomes insensitive to redshift uncertainties, and the estimator returns zero redshift correction as the best-fit value, with an uncertainty entirely determined by the observational redshift covariance. Therefore, in the limit $z \rightarrow \infty$, the data provide no additional information beyond the redshift prior, although the estimate remains statistically consistent with the true value. \\
\vskip 2pt \noindent
{\bf Case-2:} Intermediate $z$ 
\vskip 2pt \noindent At intermediate redshifts, the behaviour of the estimator reflects an interplay between the precision of the supernova magnitude measurements, the sensitivity of the likelihood to redshift perturbations, the magnitude of the peculiar velocity, and the redshift prior. In the regime where $\mathbf{M'}^T \mathbf{C}_m^{-1} \mathbf{M'} \gg \mathbf{C}_z^{-1}$, the magnitude data dominate over the prior, and the posterior uncertainty is determined primarily by the supernova magnitude covariance, i.e. $\mathbf{\Sigma}_{z^*} \simeq \left[\mathbf{M'}^T \mathbf{C}_m^{-1} \mathbf{M'}\right]^{-1}$. In this limit, the inferred true redshift is driven mainly by consistency with the magnitude–redshift relation, and the estimator effectively extracts peculiar velocity information from the data.\\ \vskip 2pt \noindent
{\bf Case-3:} Low $z$\\ 
At very low redshifts, the data term dominates the prior, since it includes the derivative of the magnitude. As $z \rightarrow0$, $m'(z)\rightarrow \infty$, and we obtain $\mathbf{\bar{z}^*=z^o}$, and $\mathbf{\Sigma}_{\zs}=\mathbf{O}$. This implies that, in the low-redshift limit, we obtain very tight constraints, primarily determined by the precision of supernova magnitude measurements. However, in this regime, the linear approximation begins to break down, as higher-order terms in the Taylor expansion of the magnitude–redshift relation become significant. As a result, the linear estimator can yield biased estimates. 

While the MCMC-based general method exhibits the same qualitative behaviour of constraints at different redshifts, it remains free from this bias, since it does not rely on a linear expansion of the magnitude–redshift relation.

\section{Trace Plots and Convergence Issues}
\label{sec:app-B}
\begin{figure}
    \centering
    \includegraphics[scale=0.55]{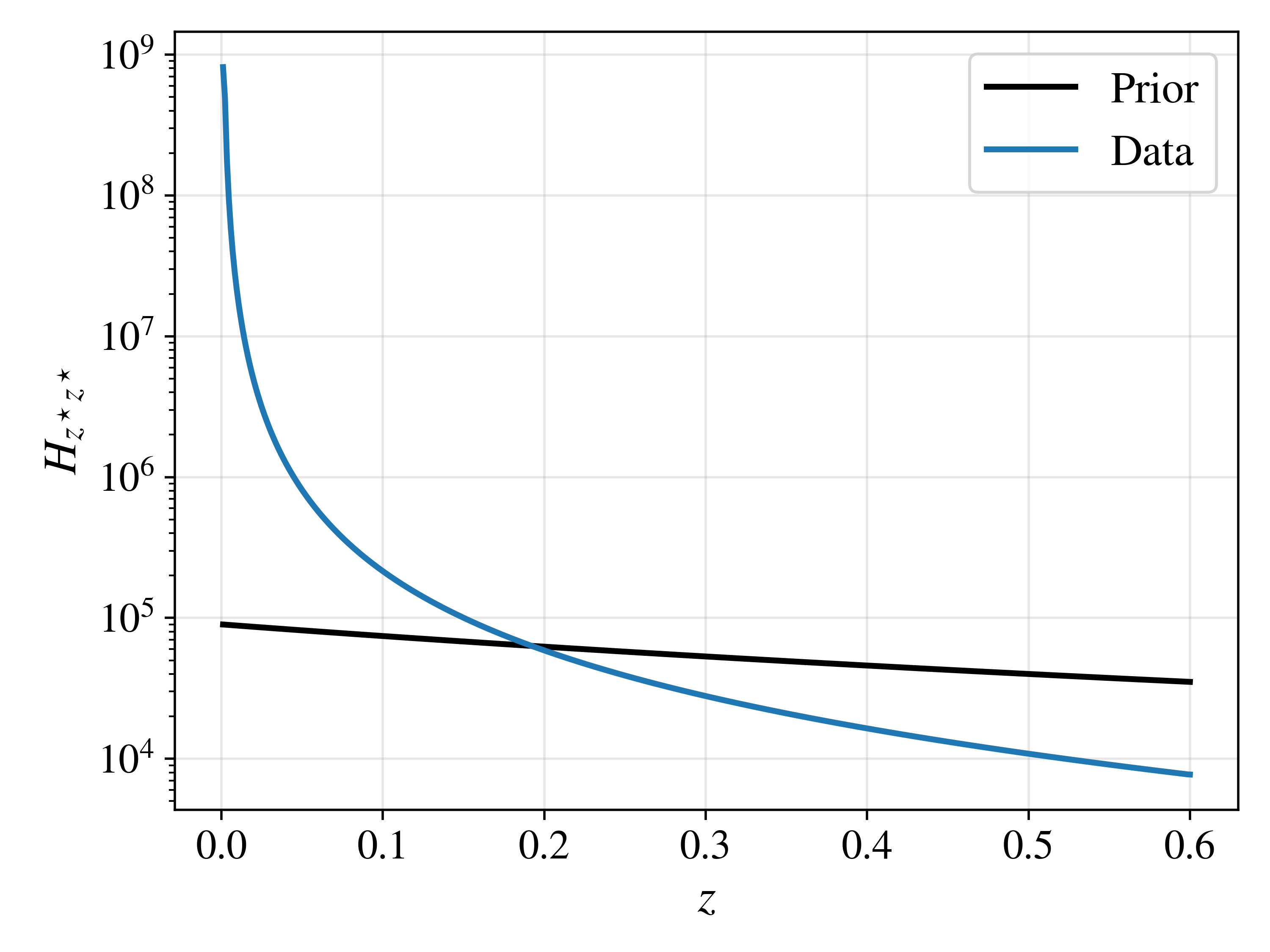}
    \caption{Comparison of the prior and data contributions to the diagonal elements of the Hessian.}
    \label{fig:Hessian}
\end{figure}

\begin{figure}
    \centering
    \includegraphics[scale=0.55]{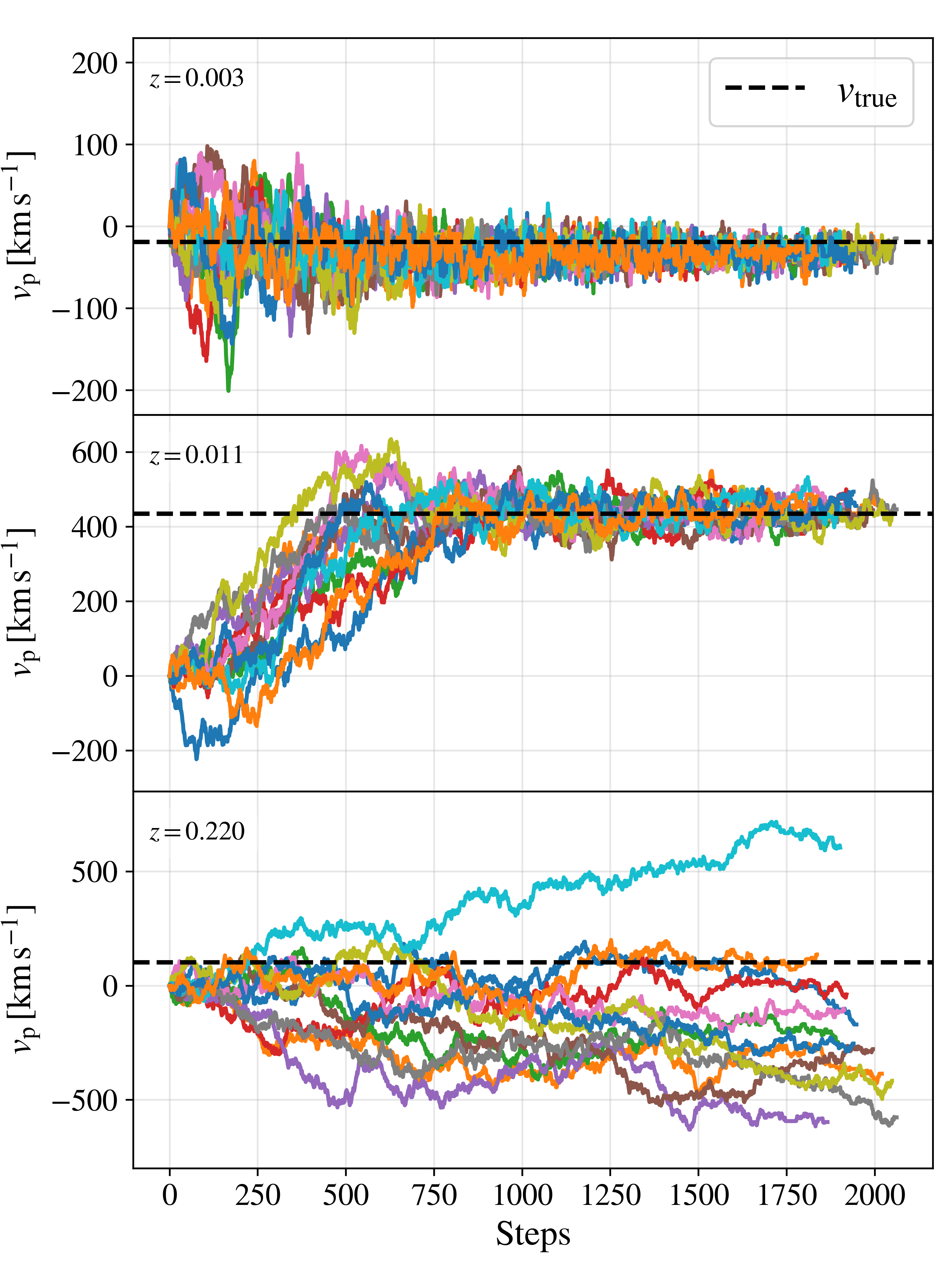}
    \caption{Trace plots illustrating the redshift dependence of MCMC convergence for the simulated data with $\sigma_m=0.02$.}
    \label{fig:trace}
\end{figure}
The issue of convergence of redshift parameters (or equivalently, peculiar velocities) discussed earlier can be easily understood with the curvature of the likelihood or the Hessian. From Eq.~(\ref{eq:lin_logpost}), the Hessian can be easily read off as the negative of the precision matrix:
\begin{equation}
    \mathbf{H} \equiv -\mathbf{\Sigma}_{\zs}^{-1}=- \left[\mathbf{C}_z^{-1} + {\mathbf{M'}^{T}} \mathbf{C}_m^{-1} \mathbf{M'}\right].
\end{equation}
It describes the total precision of the estimate, which is the sum of the prior contribution and the magnitude precision. It is more informative to compare these individual contributions to the total precision. For the simple case where the matrices $\mathbf{C}_z$ and $\mathbf{C}_m$ are diagonal, this comparison can be made by examining the diagonal elements of the Hessian,
\begin{equation}
    H_{\zs\zs} = -\left[C_{\zs\zs} + ({\mathbf{M'}^{T}} \mathbf{C}_m^{-1} \mathbf{M'})_{\zs\zs}\right].
\end{equation}
The first term represents the contribution from the prior, while the second term 
corresponds to the contribution from the data.
The redshift dependence of these two terms is shown in Figure~\ref{fig:Hessian}, where it is evident that the data dominate at low redshifts, while the prior becomes increasingly dominant at higher redshifts. The same qualitative behaviour is reflected in the trace plots of the MCMC chains for the general method shown in Figure~\ref{fig:trace}. At low redshifts, where the data dominate, the peculiar velocities are well constrained, and the chains rapidly reach a stationary behaviour. As the redshift increases, convergence deteriorates, and at high redshifts the chains wander across the full range permitted by the prior, effectively yielding no additional constraints from the data.

\bsp	
\label{lastpage}

\end{document}